\documentclass[aps,showpacs,prl,twocolumn,superscriptaddress]{revtex4}
\usepackage{graphicx}

\begin{document}
\title{Frustrated antiferromagnets at high fields:
the Bose-Einstein condensation in degenerate spectra}
\author{G. Jackeli} 
\affiliation{Institute Laue Langevin, B. P. 156, 38042
             Grenoble, France} 
\author{M. E. Zhitomirsky}
\affiliation{Commissariat \`a l'Energie Atomique,
             DSM/DRFMC/SPSMS, 38054 Grenoble, France}
              
\begin{abstract}
Quantum phase transition at the saturation field
is studied for a class of frustrated quantum antiferromagnets. 
The considered models include (i) the $J_1$-$J_2$ frustrated 
square-lattice antiferromagnet with $J_2=\frac{1}{2}J_1$  and 
(ii) the nearest-neighbor Heisenberg antiferromagnet on a 
face centered cubic lattice. 
In the fully saturated phase the magnon 
spectra for the two models have lines of degenerate minima. Transition
into partially magnetized state is treated via a mapping to
a dilute gas of  hard core bosons and by complementary spin-wave
calculations.
Momentum dependence of the exact four-point boson vertex 
removes the degeneracy of the single-particle excitation 
spectra and selects the ordering wave-vectors at
$(\pi,\pi)$ and $(\pi,0,0)$ for the two models. The asymptotic 
behavior of the magnetization curve differs significantly from 
that of conventional antiferromagnet in $d$-spatial
dimensions. We predict a unique form 
for the magnetization curve $\Delta M=S-M\simeq
\mu^{(d-1)/2}(\log\mu)^{(d-1)}$, where $\mu$ is a
distance from the quantum critical point.
\end{abstract} 
\date{December 2, 2003} 
\maketitle

Heisenberg antiferromagnets exhibit a quantum phase transition 
between a fully polarized state and a state with partial magnetization 
at the saturation field $H_c$. 
A partially magnetized state typically breaks 
spin-rotational symmetry about the field direction and has 
a long-range order of transverse spin components. Such a zero 
temperature transition belongs to the $XY$ universality class. 
Various properties of quantum antiferromagnets in the vicinity of this 
quantum critical point are well understood in terms of the Bose 
condensation of magnons below $H_c$ \cite{bb,sach}.
The antiferromagnetic wave-vector corresponds to the minimum 
in the magnon spectrum, whereas an arbitrary phase of the condensate
describes sublattice orientation in the plane 
perpendicular to the field.

The above simple picture fails, however, for so-called
frustrated antiferromagnets, which have degenerate classical 
ground states between zero and saturation 
fields \cite{frust_theory,frust_exp}. 
The excitation spectra of frustrated antiferromagnets above $H_c$ are quite
unusual. The magnon dispersion $\epsilon_{\bf q}$ 
has a continuous set of energy minima given 
by a $d'$-dimensional hypersurface embedded in the $d$-dimensional 
reciprocal space. The problem of the Bose condensation
of particles with such degenerate spectra remains unexploited to a
large extent. Numerical investigations \cite{Muetter,Honecker} of a 
specific model (model $A$ below) show  a
strong singularity of the magnetization
near the saturation field, which has not been interpreted so far.
In the present work we investigate two problems:
(i) how the quantum fluctuations remove degeneracy of the excitation
spectra and select a certain ordering wave-vector 
and (ii) how the asymptotic behavior of the magnetization curve 
close to the quantum critical point is modified 
by a presence of large phase space of soft-mode fluctuations. 
Specifically, we consider two models with 
$d'=1$ in $d=2$ and $d=3$ spatial dimensions: (model $ A$)
the $J_1$-$J_2$ antiferromagnetic Heisenberg model on a square lattice 
at its critical point $J_2=\frac{1}{2}J_1$, which has lines of minima 
at $(\pi,q)$ and $(q,\pi)$  and  (model $B$)
the nearest-neighbor Heisenberg model on a face centered cubic 
lattice (fcc) with minima at $(\pi,q,0)$ 
and equivalent lines. 

Convenient approach to deal with quantum 
antiferromagnets near the saturation field is to employ
a hard-core boson representation of spin-1/2 operators:
$S^z_i=\frac{1}{2}-b^{\dagger}_{i}b^{}_{i}$, $S^{+}_{i}=b_{i}$,
and  $S^{-}_{i}=b^{\dagger}_{i}$ with
$b^{\dagger}_{i}b^{}_{i}=0$ or 1 \cite{mm}. 
The hard-core constraint is imposed by an 
on-site repulsion $U\rightarrow\infty$.
The Heisenberg spin Hamiltonian 
$\hat{\cal H}=\sum_{\langle i,j\rangle} J_{ij} {\bf S}_i{\bf S}_j
- H \sum_i S_i^z$ is, then, transformed to
\begin{eqnarray}
\hat{\cal H}=\sum_{\bf q}
[\epsilon_{\bf q}-\mu]b^{\dagger}_{\bf q}b^{_{}}_{\bf q} 
+\frac{1}{2N}\sum_{\bf q,k, k'}V_{\bf q} b^{\dagger}_{\bf k} 
b^{\dagger}_{\bf k'} 
b^{_{}}_{\bf k'-q}b^{_{}}_{\bf k+q}\ ,
\label{H}
\end{eqnarray}
where $\epsilon_{\bf q}=\frac{1}{2}(\gamma_{\bf q}-\gamma_{\rm min})$,
$\gamma_{\bf q}=\sum_j J_{ij}e^{i{\bf q}{\bf r}_{ij}}$ 
is the Fourier transform of the exchange interaction
$J_{ij}$, $\mu=H_c-H$ is a boson chemical potential, 
and $H_c=\frac{1}{2}(\gamma_0-\gamma_{\rm min })$ 
is the saturation field.
A bare four-point boson vertex is given by 
$V_{\bf q}=U+\gamma_{\bf q}$.
The exact ground state of the Hamiltonian (\ref{H}) is the boson vacuum 
for $\mu<0$ ($H>H_c$), it corresponds to a ferromagnetic
alignment of spins.
In $d\geq 2$ the ground state with finite boson density 
$\langle b^\dagger_i b^{_{}}_i\rangle\neq 0$ at $\mu>0$ should be, generally, 
a superfluid state: $\langle b_{\bf q}\rangle\neq 0$ for a particular wave-vector
${\bf q}={\bf Q}$, such that $\epsilon_{\bf Q}-\mu$ vanishes at the saturation field. 

In the above two models
the bare spectrum $\epsilon_{\bf q}$ has lines of degenerate 
minima (see below) and the condensate wave-vector remains undetermined.
The degeneracy cannot be lifted by the bare  interaction $V_{\bf q}$
as it depends on momentum transfer only.
This is a manifestation of an infinite classical degeneracy of 
the ground state of the original
spin problem below the saturation field.
Such an `accidental' degeneracy can be removed by quantum 
fluctuations \cite{obdo}. The number of bosons vanishes at the quantum 
critical point and hence
the system becomes dilute in the limit 
$\mu\rightarrow 0^+$. 
In this case the leading order corrections to the scattering vertex
are generated by  the processes in the 
particle-particle channel. According to the standard procedure 
the bare interaction is, then, replaced by a solution of 
the integral equation for the scattering amplitude $\Gamma$ \cite{bel}.
We show that $\Gamma$ does depend on the incoming momenta and thus
lifts the degeneracy of the original problem. The ordering 
wave-vector corresponds to the smallest $\Gamma$, i.e., 
condensate occurs at the momentum at which bosons less interact
with each other in order to minimize their repulsion.

\begin{figure}[h]
\includegraphics[scale=0.4]{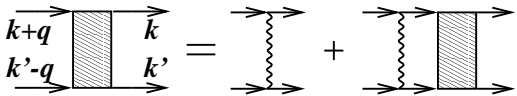}
\caption{\label{f1}Graphical representation of the integral equation for 
four-point vertex.}
\end{figure}

The Bethe-Salpeter equation for the scattering function with zero total 
frequency reads as
\begin{equation}
\Gamma_{\bf q}({\bf k},{\bf k}') = V_{\bf q}  -
\frac{1}{N}\sum_{\bf p}V_{\bf q-p}\frac{\Gamma_{\bf p}({\bf k},{\bf k}')}
{\epsilon_{{\bf k}+{\bf p}} + \epsilon_{{\bf k}'-{\bf p}}}
\label{G}
\end{equation}
and is graphically represented in Fig.~\ref{f1}.
In the limit $U\rightarrow\infty$, Eq.~(\ref{G}) is reduced 
to a system
\begin{eqnarray}
&&\Gamma_{\bf q}({\bf k},{\bf k}') = \gamma_{\bf q}+
\langle\Gamma\rangle
-\frac{1}{N}\sum_{\bf p}\gamma_{\bf q-p} 
\frac{\Gamma_{\bf p}({\bf k},{\bf k}')}
{\epsilon_{{\bf k}+{\bf p}} + \epsilon_{{\bf k}'-{\bf p}}} \ ,
\nonumber\\
&&\frac{1}{N}\sum_{\bf p}\frac{\Gamma_{\bf p}({\bf k},{\bf k}')}
{\epsilon_{{\bf k}+{\bf p}} + \epsilon_{{\bf k}'-{\bf p}}} = 1 \ ,
\label{G1}
\end{eqnarray}
where $\langle\Gamma\rangle=(1/N)\sum_{\bf q} 
\Gamma_{\bf q}({\bf k},{\bf k}')$ and the identity 
$\langle\gamma\rangle=J_{ii}=0$ has been used. 
By expanding $\Gamma_{\bf q}({\bf k},{\bf k}')$ 
in lattice harmonics of the wave-vector $\bf q$, 
the integral equations (\ref{G1}) are transformed to a system of algebraic 
equations.
Since, only a few harmonics appear in such an expansion of 
$\Gamma_{\bf q}({\bf k},{\bf k}')$,
essentially those
which are present in $\gamma_{\bf q}$, 
the resulting algebraic system
can be easily solved analytically. We shall now describe 
further details for the two models separately.

{\it Model A}: 
Frustrated antiferromagnet on a square lattice with the nearest-neighbor 
exchange constant $J_1\equiv 1$ and the diagonal coupling 
of strength $J_2=\frac{1}{2}J_1$ has 
infinitely many classical ground states for $0<H<H_c$ 
\cite{Chandra,mzh00}.
The magnon spectrum in the ferromagnetic phase at $H=H_c$ 
is $\epsilon_{\bf q}=(1+\cos q_x)(1+\cos q_y)$.
The magnon energy has lines of 
minima spanned by wave-vectors  $\{{\bf q}^*\}$: $(\pi,q)$ and $(q,\pi)$. 
Besides the well-known singularity related to vanishing of 
a scattering amplitude for two quantum particles 
in two dimensions (2d) \cite{sh}, the present 
1d type degeneracy leads to an extra infrared divergence
in the kernel of the integral equation 
when the external momenta are fixed to ${\bf k},{\bf k}'={\bf q}^*$.
Away from the critical point $\mu>0$ the singularity can be cured by
introducing an infrared cut-off for single-particle energies defined by 
$\epsilon_{\bf k,k'} \geq \mu$ \cite{sh}.
The physics behind this regularization procedure is as follows:
at the energy scale of the order of $\mu$ the interaction effects become
important and modify the form of the bare spectrum.
As it follows, an infrared behavior of the excitation spectrum 
is sound-like and, hence, removes in a self-consistent way
the divergence of the kernel below $H_c$.

The condensate wave-vector is chosen 
by considering the scattering amplitude 
at zero momentum transfer and external momenta set to 
${\bf q}^*$. The two main candidate states are
the N\'eel state with ${\bf Q}=(\pi,\pi)$, which is stable
for a frustrated square lattice antiferromagnet with weaker
diagonal bonds $J_2<\frac{1}{2}J_1$, and the columnar state 
ordered on $(\pi,0)$ [$(0,\pi)$], which is stable
for stronger diagonal bonds $J_2>\frac{1}{2}J_1$ \cite{Chandra}. 
Instead of presenting a rather lengthy solution
of the algebraic system we obtain an approximate but physically
transparent solution by neglecting the momentum-transfer 
dependence of $\Gamma_p({\bf q}^*,{\bf q}^*)\equiv\Gamma_p({\bf q}^*)$ 
near $p=0$ under the integral in equation (\ref{G1}).
The result is 
$\Gamma_0({\bf q}^*)\approx 1/\tau({\bf q}^*)$ with
$
\tau({\bf q}^*) = 
1/N\sum_{\bf p}\epsilon_{{\bf p}}/[
\epsilon_{{\bf q}^*+{\bf p}} + \epsilon_{{\bf q}^*-{\bf p}}].
$
It is evident that
$\Gamma_{0}({\bf q}^*)$ is smallest for a ${\bf q}^*$ at which the 
kernel $\tau({\bf q}^*)$ has the strongest divergence. This singles out  
the N\'eel wave-vector ${\bf Q}=(\pi,\pi)$, which has the softest
excitations $\varepsilon_{{\bf Q}+{\bf k}}\simeq k_x^2k_y^2$.
Hence, for $\mu>0$ magnons condense at ${\bf q}^*={\bf Q}$ and 
a transverse antiferromagnetic order should be formed
below the saturation field. 
Direct estimation of the integral in the kernel $\tau({\bf q}^*)$  yields
$\Gamma_0({\bf Q})\sim  \mu^{1/2}/|\log \mu|$,
whereas away from ordering wave-vector
$\Gamma_0({\bf q}^*)\sim\mu^{1/2}$. The square root behavior of
the  four-point vertex is intrinsic to 1d systems. 
This is a consequence of the quasi 1d low-energy part of the bare spectrum
of the frustrated model.
In addition, there is an extra logarithmic
correction to the 1d behavior at the ordering wave-vector ${\bf Q}$
related to vanishing stiffness in both directions leading to a 
Van Hove type singularity in the kernel.

In the ordered state below $H_c$, the
condensate density $n_0=\langle b_{{\bf Q}}\rangle^2$ is found 
from minimization  of the ground state energy density
$e_{\text{g.s.}} =-\mu n_0+\frac{1}{2}\Gamma_0({\bf  Q})n_0^2$: 
$n_0=\mu/\Gamma_0({\bf  Q}) \simeq \mu^{1/2}|\log \mu|$,
where we have neglected the noncondensate contribution.  
The excitation spectrum  
can be obtained following the standard Bogoliubov 
scheme and replacing the bare vertex $V_{\bf q}$ with 
the full scattering amplitude: 
\begin{eqnarray}
&&\omega_{\bf q}^2 = 
\left(\epsilon_{\bf q}-\mu+\Sigma^{11}_{\bf q}\right)^2
- \left(\Sigma^{12}_{\bf q}\right)^2  \  ,
\label{spectr} 
\\
&& \Sigma^{11}_{\bf q}\! =\! n_0[\Gamma_0({\bf q},{\bf Q})+
\Gamma_{{\bf q}\!-\!{\bf Q}}({\bf q},{\bf Q})] , \ 
\Sigma^{12}_{\bf q}\!=\! 
n_0\Gamma_{{\bf q}\!-\!{\bf Q}}({\bf Q}) .
\nonumber 
\end{eqnarray}
As it is seen from Eq.(\ref{spectr}) the self-energy acquires an additional
momentum dependence thanks to the dependence of $\Gamma$
on the incoming momenta.
The magnon spectrum (\ref{spectr}) is no longer degenerate 
and has 
the unique zero-energy mode at ${\bf q}={\bf Q}$.
For ${\bf q}\rightarrow{\bf Q}$ the magnon  energy becomes
$\omega_{\bf q}\approx\sqrt{4\mu [n_0\Gamma_0({\bf q},{\bf Q})-\mu]}$.
Expanding the vertex in small $|{\bf k}|=|{\bf q}-{\bf Q}|$
we obtain an acoustic mode
$\omega_{\bf k}\simeq sk$ with the velocity
$s^2\simeq \mu/|\log\mu|$, which is smaller than   
the  velocity $s^2 \simeq \mu$ 
in nonfrustrated 2d and 3d antiferromagnets near the saturation.
Away from the ordering wave-vector, along the degeneracy lines,  
magnons acquire 
a dynamically generated gap $\Delta \sim \mu|\log\mu|$.

The magnetization of the spin system is related to the total 
density of particles $M(H) = \frac{1}{2}- (n_0+n')$,
which includes both the condensate $n_0$ and noncondensate 
$n'$ parts. The noncondensate density of bosons is given by 
$n' = 1/N\sum_{\bf q}
[\epsilon_{\bf q}-\mu+\Sigma^{11}_{\bf q}-\omega_{\bf q}]/[2 \omega_{\bf q}]$.
The largest contribution to $n'$ is determined 
by wave-vectors away from $\bf Q$, along  the 
degeneracy lines ${\bf q} \in \{{\bf q}^*\}$. The ratio of the two densities
is estimated as $n'/n_0\simeq 1/|\log\mu|^{1/2}$ 
and is logarithmically small, though  it exceeds
the condensate depletion found for 
a nondegenerate 2d Bose gas ($n'/n_0 \sim 1/|\log\mu|$) \cite{sh}. 
Near $H_c$, the magnetization curve exhibits a strong singularity
\begin{equation}
M(H) \sim \mu^{1/2} |\log\mu| \sim \sqrt{H_c-H}\: |\log(H_c-H)| \ ,
\label{singularity}
\end{equation}
which fits well to the available numerical data \cite{Muetter,Honecker}.

The above results obtained in the hard-core boson picture are valid
in the vicinity of $H_c$ and should be compared with the standard
linear spin-wave theory (LSWT), which applies for all fields
$0<H<H_c$. In the LSWT approach one selects a few classical
ground states and  calculates
energies of zero-point oscillations $E_0 = \frac{1}{2}\sum_{\bf k}
\omega_{\bf k}$, which are different for each state.
In this way we have found that the N\'eel state has lower energy than a
collinear (stripe) state in the whole range of fields. Another candidate
state for $\frac{1}{2}H_c<H<H_c$ is a four-sublattice state, which has
identical polarization in three sublattices with the fourth sublattice
compensating the net transverse magnetization.
Such a partially collinear state is a natural favorite if 
quantum effects are taken into account via effective biquadratic
exchange interaction derived in a second-order real-space perturbation
theory \cite{perturb}. LSWT shows instead 
that the N\'eel state has again a lower energy
for a frustrated square-lattice antiferromagnet. This discrepancy
is explained by a nonanalytic dependence of the ground state energy
on applied magnetic field determined by
a large number of soft (zero) modes, whereas the real-space perturbation
approach captures only analytic contributions. 
Furthermore, the N\'eel antiferromagnetic order is stable even at
$H=\frac{1}{2}H_c$, where it has a lower energy of zero-point oscillations
$E_0 = 0.694S$ than a fully collinear up-up-up-down ($uuud$)
state with $E_0 = 0.703S$. This result corrects the previous
spin-wave calculation \cite{mzh00}
 and is in a good agreement 
with the exact diagonalization studies \cite{mzh00,Honecker}, 
which show that the 1/2-magnetization plateau appears in the present
model for $0.5<J_2/J_1\alt 0.65$. In the vicinity of $H_c$,
the LSWT magnetization curve for the N\'eel state shows the same type
of singularity (\ref{singularity}) with
$M = Sh\{1+\sqrt{2(1-h)}\ln[128(1-h)/\pi^4e^2]/\pi^2S\}$,
where $h=H/H_c$.

{\it Model B}: A nearest-neighbor antiferromagnet ($J\equiv 1$) on an
fcc lattice is another frustrated model
with degenerate ground states at $0<H<H_c$.
The spectrum of magnon excitation at the saturation field is
$\epsilon_{\bf q}=2[1+\cos q_x\cos q_y+\cos q_x\cos q_z+\cos q_y\cos q_z]$ 
and has lines of zeros 
at $(\pi,q,0)$ and the cubic symmetry related lines. We then  follow 
the same scheme as for the previous model and examine the kernel for
the scattering amplitude with the new spectrum. The lowest value
of renormalized four-point boson vertex corresponds to the three
wave-vectors ${\bf Q}_i=(\pi,0,0)$, $(0,\pi,0)$, and 
$(0,0,\pi)$. The corresponding vertex is estimated as 
$\Gamma=\Gamma_0({\bf Q}_i,{\bf Q}_i)\sim 1/|\log\mu|^2$. 
It has a 2d logarithmic behavior due to a
quasi-2d form of the spectrum in the vicinity of $(\pi,q,0)$ line
with an extra logarithmic
singularity related again to a vanishing stiffness  for $k=0$: 
$\epsilon_{{\bf Q}_1+{\bf k}}\simeq k_x^2+\frac{1}{4}k_y^2k_z^2$. 
Due to a presence of three equally singular wave-vectors
one  must check now 
whether the ground state of the systems is characterized 
by a Bose condensation at a single ${\bf Q}_i$
wave-vector or at all three wave-vectors simultaneously \cite{nikuni}.
For this we write the ground state energy in
the Landau form:
\begin{eqnarray}
 E & = & -\mu\sum_{i}|\psi_i|^2 +\frac{\Gamma}{2}\sum_{i}|\psi_i|^4\\
&& \mbox{} + \sum_{i\neq j}
\left[\bar{\Gamma}|\psi_i|^2|\psi_j|^2 + \frac{1}{2} 
\tilde{\Gamma}(\psi_i^{*2}\psi_j^2 + \psi_j^{*2}\psi_i^2)\right], \nonumber
\label{GL}
\end{eqnarray}
where $\psi_i=\langle b_{{\bf Q}_{i}}\rangle$ is a complex order
parameter, 
$\bar{\Gamma}=\Gamma_0({\bf Q}_{i},{\bf Q}_{j})+\Gamma_{{\bf Q}_{i}-
{\bf Q}_{j}}({\bf Q}_{i},{\bf Q}_{j})$, and 
$\tilde{\Gamma}=\Gamma_{{\bf Q}_{i}-
{\bf Q}_{j}}({\bf Q}_{i},{\bf Q}_{i})$.
A single component phase is stabilized for a sufficiently
strong repulsion between components
$\Gamma<\bar{\Gamma}-|\tilde{\Gamma}|$, whereas in the opposite
case all three components of the Bose condensate appear
with equal weights.
Direct calculations show that
$\bar{\Gamma}\sim 1/|\log\mu|$ and $\tilde{\Gamma} \sim
1/|\log\mu|^2$.
Hence, in the vicinity of  the saturation field 
dominated by the logarithmic behavior $\bar{\Gamma}\gg\Gamma$, 
$\tilde{\Gamma}$ and the single-$\bf k$
state is energetically favorable.
The boson density and asymptotic behavior of the magnetization curve
is given by $n=1/2-M(H)\sim \mu |\log\mu|^2$. We again
find a $d-1$ like form for the magnetization
with a logarithmic correction.

Corrections beyond the leading logarithms can, however, change
the energy balance.  From explicit estimate of the prefactors, one finds
that $\bar{\Gamma}\geq\Gamma$ only 
in an extremely narrow interval near the saturation
field for $\mu \alt \pi^2 e^{-20}$. Beyond this interval
$\bar{\Gamma}<\Gamma$ and a multi-$\bf k$ state 
with a real ($\tilde{\Gamma}<0$) superposition of
all three modes with equal amplitudes is stabilized as the ground state.
Such a spin structure corresponds to a partially collinear
spin configuration described above in our discussion of the model {\it A}.
The LSWT calculations confirm the above conclusion showing that for
magnetic fields as close to the saturation field
as $\Delta H/H_c=10^{-3}$ the partially collinear state is energetically more
favorable than a single-$\bf k$ state.
In contrast to the behavior of a frustrated square-lattice 
antiferromagnet we have also found an $uuud$ configuration
for the ground state at $H=\frac{1}{2}H_c$. This state
has the zero-point oscillation energy of $E_0=1.66S$, while
a single-$\bf k$ state has $E_0=1.74S$. Thus,
a 1/2-magnetization plateau should appear on the magnetization
curve of an fcc antiferromagnet. 

Finally, we make a few remarks on finite temperature behavior.
The models $A$ and $B$  are magnetic analogs 
of the weak-crystallization model \cite{brazovskii}. 
In this  model the phonon spectrum softens for
wave-vectors lying on a sphere. The phase transition to 
a crystal state appears to be of the first order even if
a mean-field theory would predict a continuous transition. 
Such a fluctuation driven first-order transition is explained by a
singular Hartree correction generated by thermal fluctuations,
which have a large phase space
\cite{brazovskii}.

For a 3d quantum fcc antiferromagnet a finite temperature transition
to the ordered state is expected for all $H<H_c$.  
In zero magnetic field neglecting quantum fluctuations
the Hartree correction is estimated as
$\Sigma_{\text{H}}\sim\sum_{\bf q} T/[\epsilon_{\bf q}+\tau]\sim 
CT|\log\tau|^2$,  where $\tau=[T-T_{\text{MF}}]/T_{\text{MF}}$ 
is a distance from a mean-field critical point 
$T_{\text{MF}}$. 
The self-consistent equation for the excitation gap becomes 
$\Delta=\tau+CT|\log\Delta|^2$. 
The point of absolute instability $\Delta=0$ cannot be reached 
at any finite $T$
indicating a first-order transition. 
For classical Heisenberg and $XY$ model on an fcc lattice 
the first-order transition at $H=0$
has been confirmed by the Monte Carlo simulations \cite{diep}.  
The transition between paramagnetic and antiferromagnetic states
is, therefore, of the first order
at low $H$ and at high $T$. The zero temperature phase transition
at $H_c$ is instead of the second order. 
This opens two possible scenarios for the $H$--$T$ phase  diagram
(a) a finite temperature tricritical point 
separates low-field (high-$T$)
first-order transition line from the $XY$ transition at high-fields;
or (b) fluctuation induced first-order transition survives down to 
zero temperature and terminates at the quantum critical end point.
For the two-dimensional model $A$, similar questions can be raised 
on the interplay  between a Berezinski-Kosterlitz-Thoulees   
(BKT) transition and a fluctuation induced first-order transition.
At high temperatures (low fields) high degree of degeneracy of the spectra
may induce the first-order transition and make the BKT instability inaccessible.
Such a scenario is, for example, realized in the theory of weak
crystallization of films \cite{andrei}. While at low temperatures (high fields) 
a usual BKT transition may take place. These interesting issues require further
investigations.

{\it Summary}: We have studied a quantum 
phase transition at high magnetic fields for a class of 
frustrated antiferromagnets in $d=2$ and $d=3$,
which have degenerate excitation spectra with lines of minima.
Momentum dependence of the exact four-point boson vertex 
removes the degeneracy and selects the ordering wave-vector 
$\bf Q$. The new spectra are sound-like near $\bf Q$ and
acquire dynamically generated gaps away from it. The asymptotic 
behavior of the magnetization curve shows the same singularity as 
a nonfrustrated model in $d-1$ dimension with an additional
logarithmic correction. The developed scheme can 
be applied to other frustrated  antiferromagnets
near the saturation and to singlet ground state systems with 
{\it degenerate} gapped triplet excitations, as, e.g., 
SrCu$_2$(BO$_3$)$_2$ \cite{srcubo} or
Cs$_3$Cr$_2$Br$_9$ \cite{cscrbr},
near the triplet condensation field $H_{c1}$.
In addition, recent experimental progress on the Bose condensation 
of alkali atoms on optical lattices \cite{greiner}
opens a new possibility for experimental study of condensate phenomena on
frustrated lattices,
the systems to which our results also apply.

We would like to thank A. Honecker and  E. I. Kats 
for interesting discussions. MEZ acknowledges financial
support from the 21st Century COE program of Kyoto University 
during his stay at the Yukawa Institute for Theoretical Physics.



\begin{thebibliography}{99}
\bibitem{bb} E.G. Batyev and L.S. Braginskii,
Zh. \'Eksp. Teor. Fiz. {\bf 87}, 1361 (1984) 
[Sov. Phys. JETP {\bf 60}, 781 (1984)].

\bibitem{sach} S. Sachdev, T. Senthil, and R. Shankar,
Phys. Rev. B {\bf 50}, 258 (1994). 

\bibitem{frust_theory}
 Magnetic systems with competing interactions, edited by H.T. Diep
 (World Scientific, Singapore, 1994).

\bibitem{frust_exp}
 C. Broholm {\it et al\/}.,
 Phys. Rev. Lett. {\bf 65}, 3173 (1990);
 P. Schiffer {\it et al}., Phys. Rev. Lett. {\bf 73}, 2500 (1994);
 A.P. Ramirez {\it et al\/}., Phys. Rev. Lett. {\bf 89}, 067202 (2002). 

\bibitem{Muetter}
 M.S. Yang and K.-H. M\"utter, Z. Phys. B {\bf 104}, 117 (1997).

\bibitem{Honecker}
A. Honecker, Can. J. Phys. {\bf 79}, 1557 (2001). 

\bibitem{mm}T. Matsubara and H. Matsuda, Prog. Theor. Phys. {\bf 16}, 569 
(1956). 

\bibitem{obdo}
 E.F. Shender, Sov. Phys. JETP {\bf 56}, 178 (1982);
 C.L. Henley, Phys. Rev. Lett. {\bf 62}, 2056 (1989).

\bibitem{bel} S.T. Beliaev, Zh. \'Eksp. Teor. Fiz. {\bf 34}, 433 (1958) 
[Sov. Phys. JETP {\bf 7}, 299 (1958)].

\bibitem{Chandra}
 P. Chandra and B. Doucot, Phys. Rev. B {\bf 38}, 9335 (1988).

\bibitem{mzh00}
 M.E. Zhitomirsky,  A. Honecker and O.A. Petrenko,
 Phys. Rev. Lett. {\bf 85}, 3269 (2000).

\bibitem{sh} 
M. Schick, Phys. Rev. A {\bf 3}, 1067 (1971); 
D.S. Fisher and P.C. Hohenberg,
Phys. Rev. B {\bf 37}, 4936 (1988).

\bibitem{perturb}
M.T. Heinil\"a and A.S. Oja, Phys. Rev. B {\bf 48}, 7227 (1993). 
\bibitem{nikuni}
T. Nikuni and H. Shiba, J. Phys. Soc. Jpn. {\bf 64}, 3471 (1995).
\bibitem{brazovskii} S.A. Brazovskii, Zh. \'Eksp. Teor. Fiz. {\bf 68}, 
175 (1975) 
[Sov. Phys. JETP {\bf 41}, 85 (1975)].

\bibitem{diep} H.T. Diep and H. Kawamura, Phys. Rev. B {\bf 40}, 7019 (1989).

\bibitem{andrei} V.V. Lebedev and A.R. Muratov, Fiz. Tverd. Tela {\bf 32}, 837 (1990) 
[Sov. Phys. Solid State {\bf 32}, 493 (1990)].

\bibitem{srcubo}
H. Kageyama {\it et al\/}., 
Phys. Rev. Lett. {\bf 82}, 3168 (1999).

\bibitem{cscrbr}
B. Leuenberger {\it et al\/}., 
Phys. Rev. B {\bf 31}, 597 (1985).

\bibitem{greiner} see, e.g., 
M. Greiner {\it et al\/}., Nature {\bf 415}, 39 (2002). 


\end{thebibliography}
\end{document}